\documentclass[a4paper,11pt]{article}
\pdfoutput=1 

\usepackage{jcappub} 

\usepackage[T1]{fontenc} 

\usepackage{graphicx}
\usepackage{color}

\usepackage{bm}

\usepackage{hyperref}
\usepackage{epsfig}

\newcommand{\itp}{\it Planck}
\newcommand{\mpl}{M_{pl}}

\title{\boldmath Inflation in the Generalized Inverse Power Law Scenario}

\author[a]{Zhun Lu}
\affiliation[a]{Department of Physics, Southeast University, Nanjing 211189, China }
\emailAdd{zhunlu@seu.edu.cn}

\abstract{We propose a single field inflationary model by generalizing the inverse power law potential from the intermediate model.
We study the implication of our model on the primordial anisotropy of cosmological microwave background radiation.
Specifically, we apply the slow-roll approximation to calculate the scalar spectral tilt $n_s$ and the tensor-to-scalar ratio $r$.
The results are compared with the recent data measured by the {\itp} satellite. We found that by choosing proper values for the parameters, our model can well describe the {\it Planck} data.
 }

\begin{document}
\maketitle
\flushbottom

\section{Introduction}
Cosmological inflation~\cite{Starobinsky:1980te,Guth:1980zm,Albrecht:1982wi,Linde:1981mu} has been recognized as a compelling attempt to solve several puzzles of
standard Big Bang cosmology, such as the flatness, horizon and monopole problems. In a modern view, the most important property of inflation is its ability to generate primordial anisotropy in the early universe~\cite{Linde:1990book,Liddle:2000book,Lyth:2009book}, which is the necessary initial condition leading to the structure formation.
Recently the observation on the anisotropy of cosmological microwave background (CMB) radiation by the {\itp} satellite~\cite{Ade:2013uln} provides unprecedent precision data on the primordial density fluctuation.
After combined with WMAP large-angle polarization data, the best fit of the scaler spectral index obtained by {\it Planck} is
$$n_s = 0.9603 \pm 0.0073, $$
Furthermore, {\it Planck} reveals the tensor-to-scalar ratio $r<0.11$ at the 95\% CL.
The {\itp} data indicate that exact scale invariance is ruled out at over 5$\sigma$, as well as that the gravitational wave contribution is small compared to the scalar contribution.
These measurements set stringent constraints on the parameter space of various inflationary models, which can be discriminated by the {\itp} data.
Strikingly, it is found that the earliest inflationary model, the $R^2$ inflation~\cite{Starobinsky:1980te,Starobinsky:1983zz}, is fully consistent with the {\it Planck} data
(see also its connection to the conformal invariant supergravity theory in Ref.~\cite{Kallosh:2013lkr}).
However, some well motivated inflationary models are found to be in conflict with the {\itp} data.
A renowned example is the original chaotic inflation~\cite{Linde:1983gd}, which is disfavored at 95\% CL.
Different approaches have been proposed to reconcile the models with the {\itp} data, such as introducing non-minimal coupling to
gravity~\cite{Kallosh:2013pby}, or choosing the non-Bunch-Davis vacuum~\cite{Ashoorioon:2013eia}.

In this work, we consider an alternative single field inflationary model, motivated by the intermediate model~\cite{Barrow:1990vx,Muslimov:1990be,Barrow:1993zq} which contains only the inverse power law term:
\begin{equation}
V(\phi)=\Lambda^4\left(M_{pl}\over \phi\right)^\beta,
\end{equation}
with $\beta>0$.
A notable feature of intermediate model is that it leads to exact solution to the equation of motion.
Like many other models, the inverse power law potential is disfavored by the {\itp} data, being outside the joint 95\% CL contour of $n_s$ and $r$ for any $\beta$.
However, as we will show blow, after slightly modifying the original model,
we find that the new form of potential, as a generalization of the inverse power law potential, can well describe the {\itp} $n_s$ and $r$ data.

\section{Generalizing the inverse power law potential}
Inflation is driven by the material with the unusual property of a negative pressure. This can be realized by a homogeneous scalar field $\phi=\phi(t)$, the inflaton,  whose evolution is governed by the equation of motion
\begin{equation}
\ddot{\phi}+3H\dot{\phi}+V^\prime=0, \label{eq:eom}
\end{equation}
and the Friedmann equation
\begin{equation}
H^2 ={1\over 3 \mpl^2} \left(V(\phi)+{1\over 2} \dot \phi^2\right),\label{eq:friedmann}
\end{equation}
where $\mpl=1/\sqrt{8\pi G}$ is the reduced Planck mass,  $V(\phi)$ is
the potential of the scalar field, and the prime denotes the derivative with respect to the scalar field $\phi$.

With quantum fluctuation, the scalar field is perturbed around its homogenous background
\begin{equation}
\phi(\bm{x},t) = \phi_0(t)+\delta \phi(\bm{x},t),
\end{equation}
where $\phi_0(t)$ is the homogenous part of the inflaton.
The Fourier modes of the inflaton fluctuation in the uniform curvature
gauge obey the following
evolution equation~\cite{Mukhanov:1988jd,Sasaki:1986hm}
\begin{equation}
(a\delta \phi_k)^{\prime\prime}+\left(k^2-{z^{\prime\prime}\over z}\right)(a\delta \phi_k) =0
\end{equation}
here the primes denote the derivative with respect to conformal time, $\delta \phi_k$ are the Fourier modes of $\delta\phi$ with the wavevector $\bm{k}$, and $z=a\dot{\phi}/H$.

The potential of the model we consider is sketched as
\begin{equation}
V(\phi) = V_0\left(1- \mu\left({M_{pl}\over \phi}\right)^\beta\right)^2,\label{pot1}
\end{equation}
where $\beta$ is a positive integer, and $\mu$ is a positive dimensional-less
coefficient.
Since the potential (\ref{pot1}) has inverse power terms and a constant term, we call it as the generalization of the inverse power
law potential.
At the region $M_{pl}\ll \phi$, the potential in Eq.~(\ref{pot1}) is simplified to
\begin{equation}
V(\phi) \approx V_0\left(1- 2 \mu\left({M_{pl}\over \phi}\right)^\beta\right). \label{pot2}
\end{equation}

\begin{figure}[ht!]
\centering
\vskip 0.2cm
\includegraphics[scale=0.9]{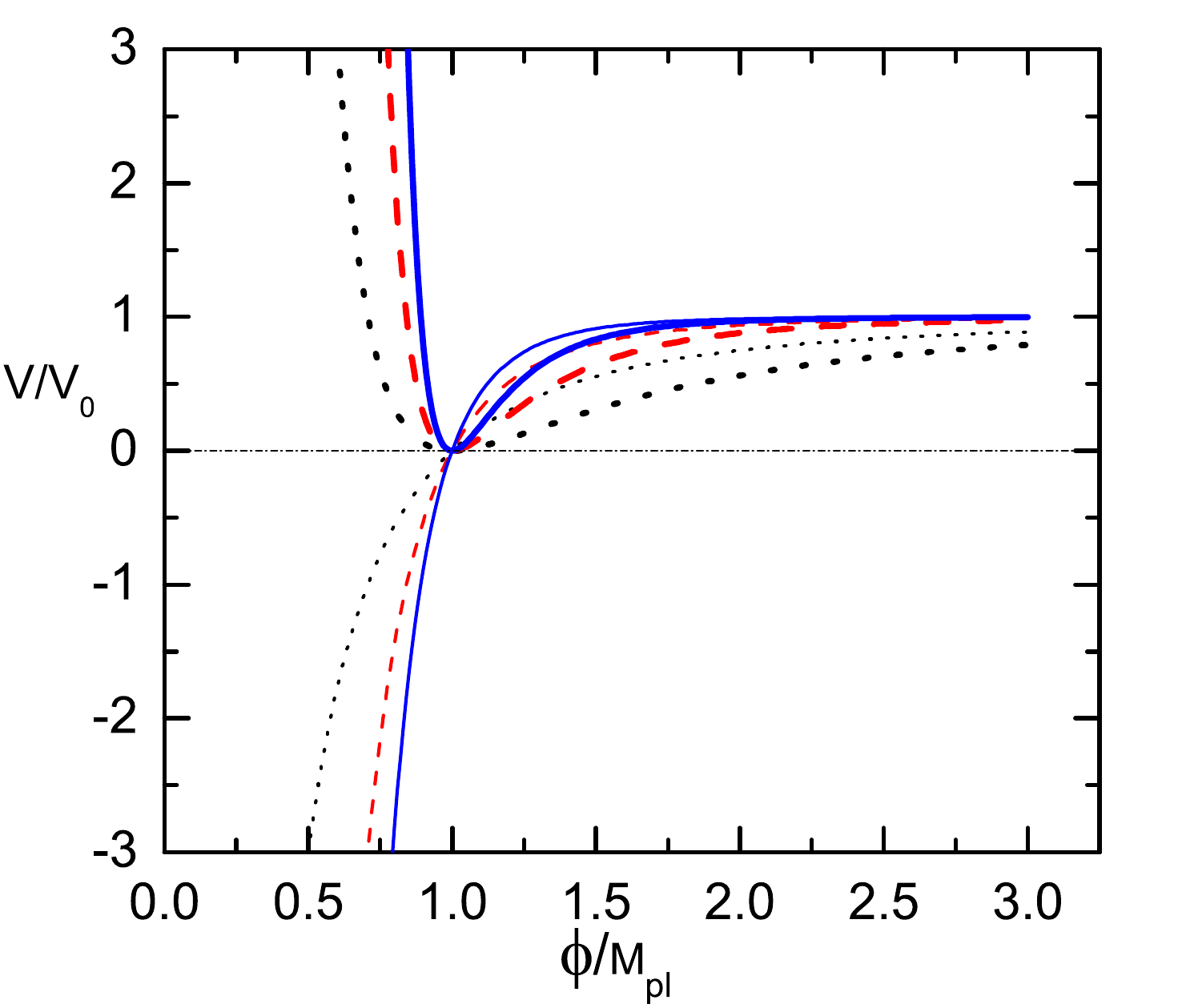}\\
  \caption{Comparisons between the potential in our model (thick lines) and that in the mutated hybrid inflationary model(thin lines). The dotted, dashed, and solid lines correspond to $\beta=2,4$ and 6, respectively. To obtain these curves we simply choose $\mu=1$ for illustration.}\label{fig:potential}
\end{figure}

In the case of $\beta=2$ or $4$, the form of the potential in (\ref{pot2}) is  similar to the potentials led from some mutated hybrid inflationary models~\cite{Stewart:1994pt,Lazarides:1996rk}. It is worthwhile to point out that the potential in (\ref{pot2}) is also motivated by the brane inflation~\cite{Dvali:2001fw,Shiu:2001sy,Kachru:2003sx}.
However, in the region where $\phi$ is comparable with or smaller than $\mu^{1\over \beta}\mpl$, the shape of the potential in (\ref{pot1}) is very different from that in (\ref{pot2}).
In Figure.~\ref{fig:potential}, we plot the shape of the potentials in (\ref{pot1})  and (\ref{pot2}), respectively.
To obtain these curves we simply choose $\mu=1$ for illustration.
In the case of (\ref{pot1}), the potential has a minimum at $\phi= \mu^{1\over \beta} \mpl$, corresponding to the vacuum expectation value (vev) of $\phi$, where the potential vanishes.
Therefore, in our model inflation starts from the region where $\phi$ is several times larger than $\mu^{1\over \beta}\mpl$, then naturally ends when the minimum is approached.
In this letter we utilize the standard technique for analyzing inflation, the so called slow-roll approximation.
In this paradigm the $\ddot{\phi}$ term in Eq.~(\ref{eq:eom}) and the $\dot{\phi}^2$ term in Eq.~(\ref{eq:friedmann}) are ignored. For this approximation to be valid, the following condition should be satisfied~\cite{Liddle:1992wi,Liddle:1993fq}
\begin{align}
\epsilon(\phi)&={\mpl^2 V^{\prime 2} \over 2 V^2}\ll 1,\\
|\eta (\phi)|& = \left|{\mpl^2 V^{\prime\prime}\over V}\right| \ll 1.
\end{align}
Using the potential given in (\ref{pot1}), we obtain the corresponding slow-roll parameters as
\begin{align}
\epsilon & ={2\mu^2\beta^2 \left({\mpl \over \phi}\right)^{2\beta+2} \over \left(1-\mu \left({M_{pl} \over \phi}\right)^\beta\right)^2 },   \\
\eta & = {2\mu^2\beta^2 \left({\mpl \over \phi}\right)^{2\beta+2} \over \left(1-\mu \left({M_{pl} \over \phi}\right)^\beta\right)^2 }
-
{2(\beta+1)\beta\mu \left({\mpl \over \phi}\right)^{\beta+2}\over 1-\mu \left({M_{pl} \over \phi}\right)^\beta }.
\end{align}
In our model, inflation starts at the region $\mu(\mpl/\phi)^\beta \ll 1$, where the slow-roll approximation is valid.
Therefore the above two parameters can be simplified to
\begin{align}
\epsilon &\approx 2\mu^2\beta^2 \left({\mpl \over \phi}\right)^{2\beta+2},
\nonumber\\
\eta & \approx
-2(1+\beta)\beta\mu \left({\mpl \over \phi}\right)^{\beta+2}.
\end{align}
We can see that the two parameters satisfy $\epsilon < |\eta| \ll 1 $ in the slow-roll limit, also in our model $\eta$ is negative.

\section{Spectral tilt and scalar-to-tensor ratio from the model}

Usually the scale dependence of the power spectra of curvature and
tensor perturbations are defined as
\begin{align}
\mathcal{P}_\mathcal{R} (k) & = A_s
\left({k\over k_*}\right)^{n_s -1+{1\over 2}
\textrm{d}n_s/d\ln k\ln(k/k_*) + \cdots}, \\
\mathcal{P}_t(k) &= A_t \left({k\over k_*}\right)^{n_t +{1\over 2}
\textrm{d}n_t/d\ln k\ln(k/k_*) + \cdots},
\end{align}
where $A_s$ ($A_t$) is the scalar (tensor) amplitude,  $n_s$ ($n_t$)
is the scalar (tensor)
spectral index, and $\textrm{d}n_s/d \ln k (\textrm{d}n_t/d \ln k) $ is the running of the scalar (tensor) spectral index, respectively.
Here we mainly focus on the scalar spectral tilt $n_s$, and the tensor-to-scalar ratio $r$ at the pivot scale.
They can be expressed by the slow-roll parameters at the leading order as:
\begin{align}
n_s -1& \approx  2\eta -6\epsilon,\\
r &={\mathcal{P}_t (k_*) \over \mathcal{P}_\mathcal{R} (k_*)}\approx 16 \epsilon.
\end{align}

The slow-roll parameters are determined by the value of the inflation field $\phi_*$ where the comoving scale $k_* = a_*H_*$ crosses the
Hubble radius for the first time.
To resolve $\phi_*$ the slow-roll approximation is usually used to calculate the   number of e-foldings after the horizon exit
\begin{align}
N_*=\int_{t_*}^{t_{\textrm{end}}} Hdt
\approx {1\over M_{pl}^2} \int_{\phi_{\textrm{end}}}^{\phi_*}
{V\over V^\prime} d\phi,\label{eq:int}
\end{align}
where the subscript ``end" denotes the end of the inflation defined by $\epsilon(\phi_{\textrm{end}}) = 1$.

Using the potential (\ref{pot1}) in our model,
we obtain the number of e-foldings as
\begin{align}
N_* & = {2\left({\phi_*\over\mpl}\right)^{\beta+2}-\mu(\beta+2)
\left({\phi_*\over\mpl}\right)^2\over 4 \mu\beta (\beta+2)}\\
& \approx{\left({\phi_*\over\mpl}\right)^{\beta+2}\over 2 \mu\beta (\beta+2)},
\end{align}
where the main contribution to $N_*$ in the integration (\ref{eq:int}) comes from the upper limit $\phi_*$.
Therefore we obtain the slow-roll parameters at the pivot scale:
\begin{align}
\epsilon(\phi_*)&= 2 \mu^{2\over \beta+2} \beta^2
\left[2N\beta(\beta+2)\right]^{-{2\beta+2\over \beta+2}},\\
\eta (\phi_*)& = -{(\beta+1)\over(\beta+2)N_*}.
\end{align}
Using the above results, the scalar spectral index $n_s$ can be expressed as
\begin{align}
n_s & \approx 1- {2(\beta+1)\over(\beta+2)N_*},
\end{align}
which is independent of parameter $\mu$.
In the limit $\beta \rightarrow \infty$, the spectral index turns to
$n_s = 1-{2\over N_*}$, mimicking the results in the $R^2$ inflationary model.
Practically, for the model being consistent with the {\itp} data, we do not need to push $\beta\rightarrow \infty$.
In the case of number of e-foldings $N_* =55$, We find that for $\beta = 4$, our model leads to $n_s=0.9698$, which is already consistent with the {\itp}+WP data ($n_s=0.9624 \pm 0.0075$) after the possible tensor component is included.

\begin{figure}[ht!]
\centering
\vskip 0.2cm
\includegraphics[scale=0.9]{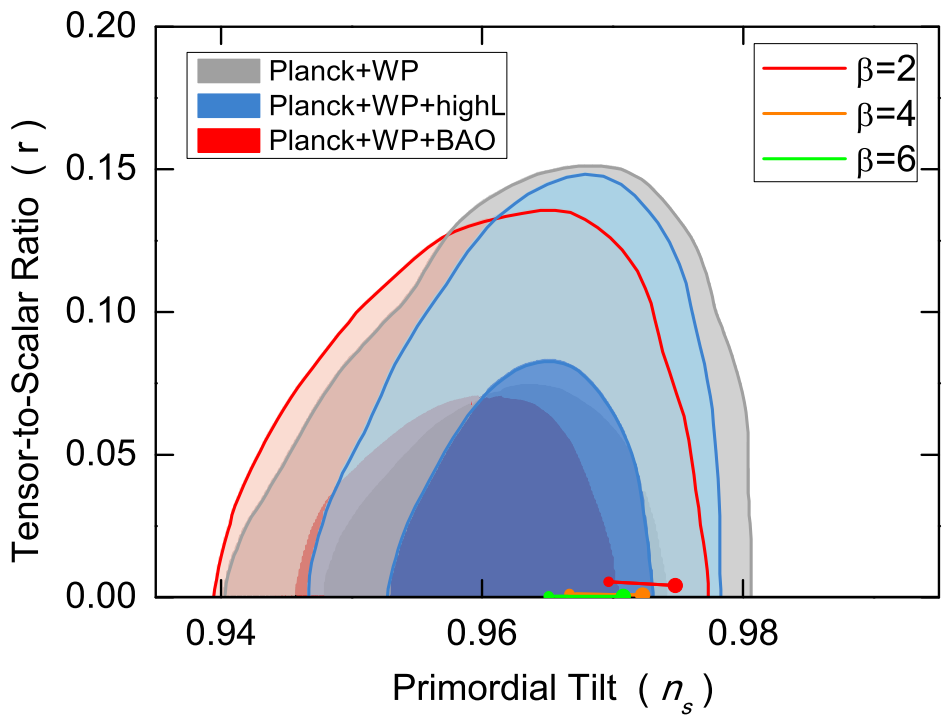}\\
  \caption{Prediction of $n_s$ and $r$ for $\beta=2,4$ and 6 compared to Marginalized joint 68\% and 95\% CL regions from {\itp}~\cite{Ade:2013uln} in combination with WP, highL and BAO data sets. The smaller and larger circles correspond to the results for $N_* = 50$ and $N_*=60$, respectively. }\label{fig:nrs}
\end{figure}

The scalar-to-tensor ratio
\begin{align}
r & = 32 \mu^{2\over \beta+2}\beta^2
\left[2N_*\beta(\beta+2)\right]^{-{2\beta+2\over \beta+2}},
\end{align}
depends both on $\beta$ and $\mu$.
However, because of the appearance of the factor $\mu^{2\over \beta+2}$, the dependence of $r$ on $\mu$ may be weak at large $\beta$.
For example, in the case of $\beta=6$, when $\mu$ varies between $10^{-4}$ and $10^4$, the factor $\mu^{2\over \beta+2}$ varies between 0.1 and 10.
Actually, in the limit $\beta \rightarrow \infty$, the dependence of $r$ on $\mu$ disappears, and $r\rightarrow 8 /(N_*^2\beta^2)\simeq 0$.
Therefore, in this limit, although the
scalar spectral index $n_s$ agrees with the $R^2$ model predictions, the tensor-to-scalar ratio $r$ does not, as the $R^2$ model predicts that $r \simeq 12/N_*^2$.
To give a robust prediction on $r$ at finite $\beta$, we still need to determine the value or the range of parameter $\mu$.
We argue that adopting $\mu\approx 1$ is a natural choice.
This corresponds that the inflaton has a vev at $\phi \approx \mpl$.
Using the COBE normalization~\cite{Bunn:1996py}
\begin{align}
{V^{1/4}\over \epsilon^{1/4}}=6.6\times 10^{16}\,\textrm{GeV},
\end{align}
we find that the potential at the pivot scale $V_*$, or equivalently, the parameter $V_0$ are constrained as
$0.45\times 10^{16} \,\textrm{GeV}< V_* \approx V_0 <0.85 \times 10^{16} \,\textrm{GeV}$
for $\beta$ between 2 and 6.

In Figure.~\ref{fig:nrs}, we present the prediction on $n_s$ and $r$ from our model for $\beta=2$ (red line), 4 (orange line) and 6 (blue line).
The smaller and larger circles correspond to $N_*=50$ and $60$, respectively.
The prediction is compared to the marginalized joint 68\% and 95\% CL regions from {\itp} in combination with WP, highL or BAO data sets.
From the figure one can see that our model predictions for $\beta=2, 4$ and $6$ are inside the 95\% CL regions of all data sets.
Especially, for $\beta=6$, the result is well consistent with the {\itp} constraints.
Numerical study shows that at the pivot scale, the size of the inflationary field is $\phi_* \approx 5.5 \mpl$, $3.7\mpl$ and $2.9\mpl$ for $\beta=2$, 4 and $6$, respectively.

\begin{table}[t]
\centering
\begin{tabular}{c||c}
\hline
  $\beta=4$ & $\beta=6$ \\
\hline\hline
\begin{tabular}{c|c|c|c}
  $\mu$ & 1 & $10^2$ & $10^3$ \\
  \hline
  $n_s $ & ~0.9698~ & ~0.9700~  &~ 0.9700 ~ \\
  \hline
   $r$ & ~0.0010~ &~ 0.0044~  & ~0.0089~ \\
  \hline
\end{tabular} & \begin{tabular}{c|c|c|c}
 $\mu$ & 1 & $10^2$ & $10^4$ \\
  \hline
  $n_s $ & ~0.9682~ & ~0.9684~  &~ 0.9686 ~ \\
  \hline
   $r$ & ~0.0004~ &~ 0.0011~  & ~0.0033~ \\
  \hline
\end{tabular}
\end{tabular}
\caption{\label{tab:nsr} Numerical results for $n_s$ and $r$ with $\mu$ varied from 1 to $10^3$ (for $\beta=4$) or 1 to $10^4$ (for $\beta=6$), The number of e-foldings after the horizon exit chosen in this calculation is $N_*=55$. }
\end{table}

Our model predicts rather small gravitational wave contribution for $\mu=1$ at large $\beta$.
In order to figure out if gravitational wave contribution is detectable at large $\beta$, we study the dependence of $r$ on the parameter $\mu$.
We vary the value of $\mu$ for fixed $N_*=55$ and show the corresponding results in Table.~\ref{tab:nsr}.
We find that as $\mu$ increases from 1 to $10^3$ (for $\beta=4$ ) or $10^4$ (for $\beta=6$), the tensor-to-scalar ratio $r$ increases almost one order of magnitude.
Thus the gravitational wave contribution might be detected at the
region $\mu$ is several order of magnitude larger than unity for $\beta=6$.
We refrain to present the result at larger $\mu$ region, because a  larger $\mu$ will give a relative large vev $\mu^{1\over\beta}\mpl$;
therefore, the inflation should start at the region where $\phi \gg \mpl$, which is very unnatural.

Finally, we consider the extreme case when the parameter $\beta$ approaches to 0.
In this limit the potential $V$ does not depend on $\phi$, thus the derivative $V^\prime$ and $V^{\prime\prime}$ vanish.
As a consequence, the slow-roll parameters are zero in this case, leading to $n_s=1$ and $r=0$.
Therefore, the extreme case $\beta\sim 0$ is ruled out by the {\itp} data.

\section{conclusion}
We have studied a class of inflationary models which are generalized from the inverse power law potential. We presented the model prediction on the scalar spectral tilt $n_s$ and the tensor-to-scalar ratio $r$.
We found that in the case of $\beta=6$, the model is well consistent with the {\itp} data.
Our model predicts rather small tensor-to-scalar for $\mu$ around unity.
The parameter $\mu$ is loosely bounded by the {\itp}  data, such that the varying of $\mu$ from 1 to $10^4$ leads to $r$ changing an order of magnitude (for $\beta=6$).
Future precision measurement on the tensor-to-scalar could be used to pin down the range of the model parameters.

\section*{Acknowledgements}
This work is partially supported by National Natural Science
Foundation of China (Grant No.~11005018, Grant No.~11120101004),
by SRF for ROCS from SEM, and by the Fundamental Research
Funds for the Central Universities.

\end{document}